\documentstyle[11pt,epsf,fleqn]{article}
\begin{document}
\title{Possibility of spontaneous CP violation in the nonminimal supersymmetric standard model with two neutral Higgs singlets}
\author{S.W. Ham$^{(1)}$, S.K. Oh$^{(1, 2)}$, and D. Son$^{(1)}$
\\
\\
{\it $^{(1)}$ Center for High Energy Physics, Kyungpook National University}\\
{\it Taegu 702-701, Korea} \\
{\it $^{(2)}$ Department of Physics, Konkuk University, Seoul 143-701, Korea}
\\
\\
}
\date{}
\maketitle
\begin{abstract}
A supersymmetric standard model with two Higgs doublets and two Higgs singlets is investigated if it can accommodate the possibility of spontaneous CP violation. 
Assuming the degeneracy of the scalar quark masses of the third generation,
we find that spontaneous CP violation in the Higgs secor is viable in our  
model. In the case of spontaneous CP violation, the masses of the lightest 
two neutral Higgs bosons are estimated to be 80 and 125 GeV for some parameter values in  our model, which are consistent with LEP2 data.
\end{abstract}
\vfil

\section{INTRODUCTION}

The violation of CP symmetry can be explained in the standard model [1] in terms of a complex phase in the Cabibbo-Kobayashi-Maskawa (CKM) matrix [2] for the mixing of the quark weak eigenstates.
Meanwhile, in principle CP symmetry may be violated through the mixing between the scalar Higgs boson and the pseudoscalar one in the neutral Higgs sector of the models with at least two Higgs doublets [3].
In practice in the minimal supersymmetric extension of the standard model (MSSM) [4], which has just two Higgs doublets, neither explicit nor spontaneous CP violation is found to be possible at the tree level, because the complex phase can always be eliminated by rotating the Higgs fields.

In the MSSM, explicit CP violation is viable through the radiatively corrected Higgs potential [5]. 
In explicit CP violation scenario, the presence of CP phase at the 1-loop level in the MSSM leads to significant modifications to both the Higgs boson mass and the mixing among them, as radiative corrections due to the third generation quark and scalar quark loops become important.
Recently, the MSSM searches for explicit CP violation were extended to radiative corrections due to the $W$ boson, the charged Higgs boson, and the chargino contributions [6].
The radiative corrections from those loops have been estimated to be significant to the mixing among the neutral Higgs bosons, especially if the ratio of the vacuum expectation values (VEVs) between two Higgs doublets is large [6]. 
Also CP symmetry in the MSSM can be spontaneously broken in the framework of the radiatively corrected Higgs sector of the MSSM. 
However, the spontaneous CP violation is disfavored because it leads to a very light neutral Higgs boson, which has already been ruled out by the Higgs search at the CERN $e^+ e^-$ collider LEP2 [7].

The Higgs sector of the MSSM can be extended by adding neutral Higgs singlet fields in addition to two Higgs doublets [8].
The simplest extention of the MSSM is next to the MSSM (NMSSM).
It might as well be called the (M+1)SSM [8].
The primary motivation for the (M+1)SSM is that this model can avoid the so-called $\mu$-problem in the MSSM.
Thus, while the MSSM introduces by hand the dimensional $\mu$ parameter, the (M+1)SSM can generate the corresponding quantity dynamically by means of the VEV of the neutral Higgs singlet field.

For CP violation, it has been found that, unlike the MSSM, the (M+1)SSM may at the tree level violate CP symmetry explicitly. 
By assuming the degeneracy of the scalar top quark masses in the Higgs sector of the (M+1)SSM, it is found that large explicit CP violation may be realized as the VEV of the neutral Higgs singlet approaches to the electroweak scale [9].
At the 1-loop level, in the (M+1)SSM, the effects of explicit CP violation to the neutral and charged Higgs boson masses are investigated by taking radiative corrections due to the quarks and scalar quarks of the third generation into account [10]. 
We have elsewhere extended by including the contributions of the loops of the $W$ boson, the charged Higgs boson, and the charginos to the neutral Higgs boson masses as well as the mixing among them, where the production cross sections for the neutral Higgs bosons are calculated via the Higgs-strahlung process at future $e^+e^-$ linear collisions when the lightest neutral Higgs boson is dominantly composed of the singlet field [11].

Meanwhile, it has been discovered that spontaneous CP violation cannot be invoked at the tree level in the (M+1)SSM, if the stability of the vacuum of the (M+1)SSM should be ensured [12].
This situation has not been improved by including radiative corrections arising from the degenerate scalar top quark masses [13].
More precisely, in this case, the CP violating minima are possible only when $\lambda$ is very small for $\tan \beta$ = 1, and thus two neutral Higgs bosons becomes very light.
In the Higgs sector of the (M+1)SSM, spontaneous CP violation may be realized by virtue of radiative corrections when the mass splitting effects between the scalar top quarks are taken into account [14].

Summing up, it has been discovered that spontaneous CP violation is not allowed at the tree level for either the MSSM or the (M+1)SSM.
Furthermore, these situations have not been improved at the 1-loop level by taking into account the contributions of the degenerate scalar top quark masses.
Quite naturally, we would like to extend the models further to see if CP symmetry might be violated spontaneously at the tree level or at the 1-loop level.
Hence we add one more neutral Higgs singlet field to the (M+1)SSM, to build a model of two Higgs doublets and two Higgs singlets.
It may be called the (M+2)SSM.

In this paper, we investigate whether CP symmetry can be spontaneously broken in the Higgs potential of the (M+2)SSM. 
To explore the implication of spontaneous CP violation on the (M+2)SSM in more detail, we concentrate on the case in which the scalar quarks of the third generation become degenerate in their masses.
We investigate whether spontaneous CP violation can be realized for the Higgs sector of the (M+2)SSM in the case with radiative effects of the degenerate scalar quark masses.
We would not consider explicit CP violation scenario in the Higgs potential of the (M+2)SSM.
Thus, we assume that all the relevant parameters in its Higgs potential are real.
On the other hand, in order to realize spontaneous CP violation in the Higgs potential, the VEVs of the two Higgs doublets as well as of the two Higgs singlets are assumed to be complex.

We describe the tree-level Higgs potential of the (M+2)SSM in spontaneous CP violation scenario.
Since in supersymmetric models the contribution of radiative corrections is important for studying the Higgs sector, we calculate the neutral Higgs boson masses as well as the mixing among them by using the 1-loop effective potential with the contributions of the degenerate scalar quark masses in the context of the (M+2)SSM. 
We find that spontaneous CP violation is possible for the wide parameter space of the considered Higgs potential of the (M+2)SSM without contradicting the phenomenological constraints set by the experimental data of LEP2.

Our paper is organized as follows.
In the next section, we present the Higgs potential of the (M+2)SSM at the tree level and the 1-loop effective potential with the contribution of the degenerate scalar quark masses.
In the third section, we calculate the (M+2)SSM Higgs boson masses in spontaneous CP violation sceanrio by considering the negative experimental results on the Higgs search at LEP2.
Conclusions are given in the last section.
Appexdix supplies the formulae for the elements of the neutral Higgs boson mass matrix.

\section{THE (M+2)SSM HIGGS POTENTIAL}

The Higgs sector of the (M+2)SSM contains two Higgs  doublets and two Higgs singlets.
Let us introduce them: $H_1^T$ =  ($H_1^0$, $H_1^-$)  and $H_2^T$  = ($H_2^+$,  $H_2^0$) are the Higgs doublets, and $N_1$ and $N_2$ are two Higgs singlets.
They will yield seven neutral Higgs bosons. 
The most general form of the Higgs superpotential of the (M+2)SSM, containing the term with the quark Yukawa couplings for the third generation, is given by
\begin{eqnarray}
{\cal W} & = & h_b Q H_1 b_R^c + h_t Q H_2 t_R^c + \lambda_1 H_1^T \epsilon H_2 N_1 + \lambda_2 H_1^T \epsilon H_2 N_2 \cr
& &\mbox{} - {k_{11} \over 3} N_1^3 - {k_{22} \over 3} N_2^3 - {k_{12} \over 3} N_1 N_2^2 - {k_{21} \over 3} N_1^2 N_2  \nonumber
\end{eqnarray}
where $\epsilon$ is the usual antisymmetric 2 $\times$ 2 matrix with $\epsilon_{12} = - \epsilon_{21}$ = 1, and $\lambda_i$ and $k_{ij}$ ($i, j$ = 1, \ 2) are dimensionless coupling constants. 
The superfield $Q^T$ = ($t_L, b_L$) consists of the left-handed quarks of the third generation. The superfields $t_R^c$ and $b_R^c$ denote the charge conjugate of the right-handed quarks of the third generation.
Note that all quark and lepton Yukawa couplings except for those of the third generation quarks are neglected.
The presence of the cubic terms of the Higgs singlets in the above superpotential breaks a global U(1) Peccei-Quinn symmetry which leads to a pseudo-Goldstone boson at the tree level [15].  

From the above superpotential, the relevant tree-level Higgs potential in the (M+2)SSM is obtained as
\begin{equation}
        V^0 =  V_D + V_F + V_{\rm S}  \ ,
\end{equation}
where $D$ terms, $F$ terms, and soft terms are given as follows:
\begin{eqnarray}
V_D & = & {g_2^2\over 8}
     (H_1^\dagger\hat\sigma H_1 + H_2^\dagger\hat\sigma H_2)^2
      + {g_1^2\over 8}(|H_2|^2-|H_1|^2)^2 \ , \cr
V_F & = & |\lambda_1 N_1 + \lambda_2 N_2|^2 (|H_1|^2+|H_2|^2) \cr
& &\mbox{} + | \lambda_1 H_1^T \epsilon H_2 - k_{11} N_1^2
     - {1 \over 3} k_{22} N_2^2 - {2 \over 3} k_{21} N_1 N_2 |^2   \cr
& &\mbox{} + | \lambda_2 H_1^T \epsilon H_2 - {1 \over 3} k_{11} N_1^2
     - k_{22} N_2^2 - {2 \over 3} k_{12} N_1 N_2 |^2          \ , \\
V_{\rm S} &=& m_{H_1}^2|H_1|^2 + m_{H_2}^2 |H_2|^2
     + m_{N_1}^2 |N_1|^2 + m_{N_2}^2 |N_2|^2 \cr
& &\mbox{}  - (\lambda_1 A_{\lambda_1} H_1^T \epsilon H_2 N_1
     + \lambda_2 A_{\lambda_2} H_1^T \epsilon H_2 N_2 + {\rm H.c.} )  \cr
& &\mbox{} - \left ( {k_{11} \over 3} A_{k_{11}} N_1^3
   + {k_{22} \over 3} A_{k_{22}} N_2^3
   + {k_{12} \over 3} A_{k_{12}} N_1 N_2^2
   + {k_{21} \over 3} A_{k_{21}} N_1^2 N_2 + {\rm H.c.} \right ) \ , \nonumber
\end{eqnarray}
where $g_1$ and $g_2$ are the U(1) and SU(2) gauge coupling costants, respectively, $\hat\sigma = (\sigma^1,\sigma^2,\sigma^3)$ are the Pauli matrices, $m_{H_i}^2$ and  $m_{N_i}^2$ ($i$  = 1, 2) in the soft terms are the soft supersymmetry (SUSY) breaking masses, and $A_{\lambda_i}$ and $A_{k_{i j}}$ ($i, j$  = 1, 2) are the trilinear soft SUSY breaking parameters with mass dimension. 
In supersymmetric models, the most important contributions to the 1-loop effective potential come from the loops of the quarks and scalar quarks of the third generation.
Neglecting the mixing in the scalar quark mass matrix [13, 16], the 1-loop effective potential is given as [17]
\begin{eqnarray}
V^1 & = & {3 \over 16 \pi^2} \left \{ (h_t^2 |H_2|^2 + m_{\tilde t}^2)^2
   \log \left ({h_t^2 |H_2|^2 + m_{\tilde t}^2 \over \Lambda^2} \right )   
   -  h_t^4 |H_2|^4 \log \left ({h_t^2 |H_2|^2 \over \Lambda^2} \right)
  \right \}  \cr
& &\mbox{} + {3 \over 16 \pi^2} \left \{ (h_b^2 |H_1|^2 + m_{\tilde b}^2)^2
   \log \left ({h_b^2 |H_1|^2 + m_{\tilde b}^2 \over \Lambda^2} \right )   
   -  h_b^4 |H_1|^4 \log \left ({h_b^2 |H_1|^2 \over \Lambda^2} \right)
  \right \}
\ ,
\end{eqnarray}
where $\Lambda$ is the renormalization scale in the modified minimal substraction (${\overline {\rm MS}}$) scheme and the soft SUSY breaking masses satisfy $m_{{\tilde t}_L}^2$ = $m_{{\tilde t}_R}^2$ $\equiv$ $m_{\tilde t}^2$ $\gg$ $m_t^2$ and  $m_{{\tilde b}_L}^2$ = $m_{{\tilde b}_R}^2$ $\equiv$ $m_{\tilde b}^2$ $\gg$ $m_b^2$.
Thus the full Higgs potential $V$ is composed of $V^0$ and $V^1$.

In this model, the charged Higgs sector automatically conserves CP. 
Under reasonable assumptions, either explicit CP violation or spontaneous one can occur in the neutral Higgs sector through the mixing of the scalar and pseudoscalar Higgs bosons. 
We prohibit the possibility of explicit CP violation in the above Higgs sector.
Thus all the dimensionless coupling constants and the trilinear soft SUSY breaking parameters in the Higgs potential are assumed to be real.  
For simplicity, we assume here that some of the parameters satisfy the following relations: 
$\lambda_1 = \lambda_2$, $A_{\lambda_1} = A_{\lambda_2}$, $k_{11} = k_{22}$,  
$k_{12} = k_{21}$, $A_{k_{11}} = A_{k_{22}}$ and $A_{k_{12}} = A_{k_{21}}$. 
Thus at the tree level $m_{N_1}^2$ and $m_{N_2}^2$ have same values.
Of course the two soft SUSY breaking masses can take as different values. 
At the 1-loop level, the values of $m_{N_2}^2$ deviate from those of $m_{N_1}^2$ by considering radiative effects of the degenerate scalar quark masses.

Let us first consider the case of CP conservation. 
In this case, the Higgs potential should be such that the VEVs of the four neutral Higgs fields are all real: $\langle 0|H_1^0|0 \rangle = v_1$, $\langle 0|H_2^0|0 \rangle = v_2$, $\langle 0|N_1|0 \rangle = x_1$, and $\langle 0|N_2|0 \rangle  =  x_2$, after the electroweak symmetry breaking.  
As is  well known, the charged Higgs fields do not develop the VEVs because the U(1) gauge symmetry has to be guaranteed in the vacuum.
The four neutral scalar Higgs bosons can be distinguished from the three neutral pseudoscalar ones, as there is no mixing between them in the case of CP conservation. 
Among the four neutral scalar Higgs bosons, the one that has the smallest mass  is referred to as the lightest scalar Higgs boson. 
The upper bound on its mass is expressed as 
\begin{eqnarray}
& & m_Z^2 \cos 2 \beta^2 + v^2 (\lambda_1^2 + \lambda_2^2) \sin^2 2 \beta \cr
& &\mbox{} + {3 m_b^4 \over 4 \pi^2 v^2 \cos^2 \beta} \log \left (1 + {m_{\tilde b}^2 \over m_b^2} \right ) 
 + {3 m_t^4 \over 4 \pi^2 v^2 \sin^2 \beta} \log \left (1 + {m_{\tilde t}^2 \over m_t^2} \right )    \ ,
\end{eqnarray}
where the ratio of the VEVs of the two Higgs doublets is given by $\tan \beta$ and the neutral gauge boson mass is defined as  $m_Z^2$ = $(g_1^2 + g_2^2) v^2/2$ with $v$ = $\sqrt{v_1^2 + v_2^2}$ = 175 GeV.
Note that the above expression does not depend both on the VEVs of the neutral Higgs singlets and on the renormalization scale. 
In the above formula, the last two terms come from radiative corrections due to the degenerate scalar quark masses.
If CP symmetry is conserved in the Higgs sector, the tree-level upper bound on the lightest scalar Higgs boson mass depends critically on the sizes of $\lambda_i$ ($i$ = 1, 2).
Notice that the above expression is essentially equal to the upper bound on the lightest scalar Higgs boson mass of the (M+1)SSM. 
Thus, even though an additional Higgs singlet is present in Higgs sector of the (M+2)SSM, the upper bound on the lightest scalar Higgs boson mass is essentially the same as the corresponding one in the (M+1)SSM. 

Now, consider the case of spontaneous CP violation in the (M+2)SSM.
In order to accommodate spontaneous CP violation in the Higgs sector, we have to allow complex values for the VEVs of the four neutral Higgs fields as
\begin{eqnarray}
        && \langle 0|H_1^0|0 \rangle = v_1 e^{i \varphi_1}  \ , \cr
        && \langle 0|H_2^0|0 \rangle = v_2 e^{i \varphi_2} \ , \cr
        && \langle 0|N_1|0 \rangle = x_1 e^{i \varphi_3}  \ , \cr
        && \langle 0|N_2|0 \rangle = x_2 e^{i \varphi_4} \ , \nonumber
\end{eqnarray}
where we take $v_i$ and $x_i$ ($i = 1, 2$) to be real and positive. 
The conditions that the Higgs potential becomes minimum at  $v_1$, $v_2$, $x_1$, and $x_2$ yield four constraints, which can eliminate the soft SUSY breaking masses $m_{H_1}^2$, $m_{H_2}^2$, $m_{N_1}^2$ and $m_{N_2}^2$ from the potential $V$.

By redefining the overall phase of the Higgs potential, a combination of the four complex phases can be absorbed away, leaving three independent combinations  of the phases. 
They may be chosen to be
\begin{eqnarray}
        \phi & = & \varphi_1 + \varphi_2 + \varphi_3 + \varphi_4 \ , \cr
        \phi_1 & = & 3 \varphi_3 \ , \cr
        \phi_2 & = & 3 \varphi_4 \ . \nonumber
\end{eqnarray}
Note that there is no relative phase among the VEVs in $V^1$.
As the Higgs potential at the tree level evokes the spontaneous CP violation, the vacuum of the model has to be defined as a stationary point with respect to the three physical CP violating phases, $\phi$, $\phi_1$, and $\phi_2$. 
The minimum condition for the vacuum with respect to $\phi$ yields 
\begin{eqnarray}
        k_{11} & = &\mbox{} - {\displaystyle {3 A_{\lambda_1} x_1 \sin
        \left (\phi - {\phi_2 \over 3} \right) 
        + 8 k_{12} x_1 x_2 \sin (2\phi - \phi_1 -\phi_2) 
        + 3 A_{\lambda_1} x_2 \sin \left (\phi -{\phi_1 \over 3} \right)} 
        \over {\displaystyle 4 x_1^2 \sin \left (\phi - \phi_1 - {\phi_2 \over         3}\right) + 4 x_2^2 \sin \left (\phi - {\phi_1 \over 3} - \phi_2 \right        )} } \ ,
\end{eqnarray}
while the remaining two minimum conditions for a vacuum with respect to $\phi_1$ and $\phi_2$ may be arranged to yield expressions for $A_{k_{11}}$ and $A_{k_{12}}$ as
\begin{eqnarray}
        A_{k_{11}} 
        & = & {A_{\lambda_1} v^2  \lambda_1 x_2 \sin 2 \beta   \over 2 k_{11} 
        x_1^3 \sin (\phi_1)} \sin \left (\phi - {\phi_1 \over 3} \right)
        + {2 v^2 \lambda_1 x_2^2 \sin 2 \beta \over 3 x_1^3  \sin (\phi_1)}  
        \sin \left (\phi - {\phi_1 \over 3} - \phi_2 \right) \cr
        & &\mbox{} + {2  k_{12} v^2 \lambda_1  x_2 \sin 2  \beta \over k_{11} 
        x_1^2 \sin (\phi_1)} \sin (2 \phi  - \phi_1 - \phi_2) 
        +  {8 k_{12} x_2^3 \over 9  x_1^2 \sin  (\phi_1)} 
        \sin \left ({\phi_1 - \phi_2 \over 3} \right) \cr
        & &\mbox{} - {A_{k_{12}} k_{12} x_2^2 \over 3 k_{11} x_1^2 
        \sin (\phi_1)} \sin \left ({\phi_1 + 2 \phi_2 \over 3} \right)
        + {v^2 \lambda_1^2 x_2  \over k_{11} x_1^2 \sin  (\phi_1)} 
        \sin \left ({\phi_1 - \phi_2 \over 3} \right) \cr
        & &\mbox{} + {4 k_{11} x_2^2 \over 3 x_1 \sin (\phi_1)} 
        \sin \left ({2 \phi_1 - 2 \phi_2 \over 3} \right) 
        + {2 v^2 \lambda_1  \sin 2  \beta  \over x_1  
        \sin (\phi_1)} \sin \left (\phi - \phi_1 - {\phi_2 \over 3} \right) \cr
        & &\mbox{} - {2 k_{12} A_{k_{12}} x_2 \over 3 k_{11} x_1 \sin (\phi_1)}        \sin \left ({2 \phi_1 + \phi_2 \over 3} \right) 
        +  {8 k_{12} x_2 \over 9 \sin (\phi_1)} 
        \sin \left ({\phi_1 - \phi_2 \over 3} \right ) \ ,
\end{eqnarray}
and 
\begin{equation}
        A_{k_{12}} = {A_k^n \over A_k^d} \ ,
\end{equation}
where  
\begin{eqnarray}
A_k^d & = & {k_{12} x_2^2 \over  3 k_{11} x_1^2 \sin (\phi_1)}  
   \sin \left ({\phi_1 + 2 \phi_2 \over 3} \right)
   - {k_{12} x_1^2 \over  3 k_{11} x_2^2 
   \sin (\phi_2)} \sin \left ({2 \phi_1 + \phi_2 \over 3} \right) \cr
   & &\mbox{} + {2 k_{12} x_2 \over  3 k_{11} x_1 \sin (\phi_1)} 
   \sin \left ({2 \phi_1 + \phi_2 \over 3} \right) 
   - {2 k_{12} x_1 \over  3 k_{11} x_2 \sin (\phi_2)}  
   \sin \left ({\phi_1 + 2 \phi_2 \over 3} \right)   \ , \cr
   & & \cr
A_k^n & = & {v^2 \lambda_1 x_2 \sin 2 \beta \over 6 k_{11} x_1^3 
   \sin (\phi_1)} \left \{4  k_{11} x_2 
   \sin  \left (\phi -  {\phi_1 \over 3}  - \phi_2 \right)  
   + 3 A_{\lambda_1} \sin \left (\phi - {\phi_1 \over 3} \right) \right \} \cr
   & &\mbox{} - {v^2 \lambda_1 x_1 \sin 2 \beta \over 6 k_{11} x_2^3 
   \sin (\phi_2)} \left \{ 4  k_{11} x_1 
   \sin \left (\phi  - \phi_1 - {\phi_2 \over  3} \right) 
   + 3 A_{\lambda_1} \sin \left (\phi - {\phi_2 \over 3} \right ) \right \} \cr
   & &\mbox{} - {2 k_{12} x_1 \over 9 k_{11} x_2^2 \sin (\phi_2)} 
   \left \{9 v^2 \lambda_1 \sin 2 \beta  \sin (2 \phi - \phi_1 - \phi_2)  
   - 4 k_{11} x_1^2 \sin \left ({\phi_1 - \phi_2 \over 3} \right) \right \} \cr
   & &\mbox{} + {2 k_{12} x_2 \over 9 k_{11} x_1^2  \sin (\phi_1)} 
   \left \{9 v^2 \lambda_1 \sin 2 \beta  \sin (2 \phi  - \phi_1 - \phi_2)  
   + 4 k_{11} x_2^2 \sin \left ({\phi_1 - \phi_2 \over 3} \right) \right \} \cr
   & &\mbox{} - {2 \over 3 x_2 \sin (\phi_2)} \left  \{3 v^2 \lambda_1  
   \sin 2 \beta \sin \left (\phi - {\phi_1 \over 3} - \phi_2 \right ) 
   - 2 k_{11} x_1^2 \sin \left ( {2 \phi_1 - 2 \phi_2 \over 3} \right) 
   \right \} \cr
   & &\mbox{} + {2 \over 3 x_1 \sin (\phi_1)} 
   \left \{3 v^2 \lambda_1 \sin 2 \beta 
   \sin \left (\phi - \phi_1 - {\phi_2 \over 3} \right ) 
   + 2 k_{11} x_2^2 
   \sin \left ({2 \phi_1 - 2 \phi_2 \over 3} \right) \right \} \cr
   & &\mbox{} + {v^2 \lambda_1^2 x_2   \over k_{11}  x_1^2 \sin (\phi_1)}  
   \sin \left ({\phi_1 - \phi_2 \over 3} \right )
   + {v^2 \lambda_1^2 x_1  \over k_{11} x_2^2 \sin  (\phi_2)} 
   \sin \left ({\phi_1 - \phi_2 \over 3} \right ) \cr
   & &\mbox{} + {8 \over 9} k_{12} 
   \sin \left ( {\phi_1 - \phi_2 \over 3} \right ) 
   \left \{{x_1 \over \sin (\phi_2)} +  {x_2 \over \sin (\phi_1)} \right \} \ .
 \nonumber
\end{eqnarray}
In order words, the three minimum equations with respect to $\phi$, $\phi_1$, and $\phi_2$ determine $k_{11}$, $A_{k_{11}}$ and $A_{k_{12}}$.
They are thus no longer independent parameters but dependent on others as given in the above expressions, determined by the solutions of the above three minimum equations.
Consequently, the Higgs sector of the (M+2)SSM has as free parameters $\phi$, 
$\phi_1$, $\phi_2$, $\tan \beta$, $x_1$, $x_2$, $A_{\lambda_1}$, $\lambda_1$, $k_{12}$, $m_{\tilde t}^2$, and $m_{\tilde b}^2$.

In order to calculate the Higgs boson masses, let us redefine the Higgs fields of the (M+2)SSM in unitary gauge as
\begin{eqnarray}
        \begin{array}{lll}
        H_1 & = & \left [ \begin{array}{c}
          (v_1 + S_1 + i \sin \beta A)   \cr
          \sin \beta C^{+ *}
        \end{array} \right ]  \ ,  \cr   \cr
        H_2 & = & \left [ \begin{array}{c}
          \cos \beta C^+           \cr
        e^{i (\phi-\phi_1/3 - \phi_2/3)}  (v_2 + S_2 + i \cos \beta A)
          \end{array} \right ] \ ,   \cr   \cr
        N_1 & = & [ \begin{array}{c}
        e^{i \phi_1/3}  (x_1 + X_1 + i Y_1)
        \end{array} ]  \ , \cr  \cr
        N_2 & = & [ \begin{array}{c}
        e^{i \phi_2/3}  (x_2 + X_2 + i Y_2)
        \end{array} ]  \ ,
        \end{array}
\end{eqnarray}
where $C^+$ is the charged Higgs field, and in the absence of spontaneous CP violation, $S_i$ and $X_i$ ($i$ = 1, 2) are the neutral scalar fields and $A$ and $Y_i$ ($i$ = 1, 2) are the neutral pseudoscalar fields.
The masses of the neutral Higgs bosons are obtained as the eigenvalues of the $7\times7$ mass matrix in the basis of ($S_1$, $S_2$, $A$, $X_1$, $X_2$, $Y_1$, $Y_2$), whereas the mass of the charged Higgs boson is rather straightforwardly obtained by differentiating twice the Higgs potential with respect to the charged Higgs field. 
Since the Higgs sector of the (M+2)SSM is larger than that of the  (M+1)SSM by 
a neutral Higgs singlet, the charged Higgs sector of the (M+2)SSM is equal to that of the (M+1)SSM. 

The mass of the charged Higgs boson is thus obtained as
\begin{eqnarray}
        m_{H^+}^2 & = & m_W^2 - \lambda_1^2 v^2 + {2 \lambda_1
        A_{\lambda_1} \over \sin 2 \beta} \left \{x_1 \cos \left (\phi -
        {\phi_2 \over 3} \right ) + x_2 \cos \left (\phi - {\phi_1 \over
        3} \right ) \right \} \cr
        & &\mbox{} + {8 \lambda_1 k_{11} \over 3 \sin 2 \beta} 
        \left \{x_1^2 \cos \left (\phi - \phi_1 - {\phi_2 \over 3} \right ) 
   + x_2^2 \cos \left (\phi - {\phi_1 \over 3} - \phi_2 \right ) \right \} \cr
        & &\mbox{} + { 8 \lambda_1 k_{12} x_1 x_2 \over 3 \sin 2 \beta}
        \cos ( 2 \phi - \phi_1 - \phi_2)   \ ,
\end{eqnarray}
where the charged gauge boson mass is defined as $m_W^2$ = $g_2^2 v^2/2$.
The tree-level mass of the charged Higgs boson is not changed by taking into account radiative corrections due to the degeneracy of scalar quark masses.
Note that complex phases enter into the charged Higgs boson mass.
This is remarkable because spontaneous CP violation does not occur in the charged Higgs sector but in the neutral Higgs sector. 
As mentioned before, the charged Higgs sector has an intact CP symmetry since the pair of the charged Higgs bosons are CP-even. 
Hence, we might say that the effect of the spontaneous CP violation in the neutral Higgs sector is transferred or shifted to the charged Higgs sector.
The shift effect on the mass of the charged Higgs boson would vanish if CP symmetry is conserved in the neutral Higgs sector, {\it i.e.}, if $\phi$ = $\phi_1$ = $\phi_2$ = 0.
Then we have 
\[
        m_{H^+}^2  = m_W^2 - \lambda_1^2 v^2 + {2 \lambda_1
        A_{\lambda_1} (x_1 + x_2) \over \sin 2 \beta} 
        + {8 \lambda_1 k_{11} (x^2_1 + x^2_2) \over 3 \sin 2 \beta} 
        + { 8 \lambda_1 k_{12} x_1 x_2 \over 3 \sin 2 \beta}  \ ,
\]
which may further be reduced to the tree-level mass of the charged Higgs boson in the (M+1)SSM with CP  conservation, if the  relevant parameters  are set as  $\lambda$ = $\lambda_1$, $A_{\lambda}$ = $A_{\lambda_1}$, $k$ = $k_{11}$ = $k_{12}$, and $x$/2 = $x_1$ = $x_2$. 

Now, let us derive expressions for the $7\times7$ symmetric neutral Higgs mass matrix in the ($S_1$, $S_2$, $A$, $X_1$, $X_2$, $Y_1$, $Y_2$)-basis,
It is calculated by differentiating twice the Higgs potential with respect to the seven neutral Higgs fields. 
The elements of the neutral Higgs boson mass matrix, $M_{i j}$ ($i, j$ = 1 to 7) at the tree level are given in Appendix.
By considering the radiative effect of the degenerate scalar quark masses,
the only modification to the tree-level mass matrix for the neutral Higgs boson
is
\begin{eqnarray}
\delta M_{1 1} & = & {3 m_b^4 \over 4 \pi^2 v^2 \cos^2 \beta} \log \left (1 + {m_{\tilde b}^2 \over m_b^2} \right ) \ , \cr
\delta M_{2 2} & = & {3 m_t^4 \over 4 \pi^2 v^2 \sin^2 \beta} \log \left (1 + {m_{\tilde t}^2 \over m_t^2} \right )    \ .
\end{eqnarray}

In the CP conserving limit, when $\phi$ = $\phi_1$ = $\phi_2$ = 0, the mixing elements among the scalar components and the pseudoscalar ones in the neutral Higgs boson mass matrix, $M_{ij}$ ($i$ = 1, 2, 4, 5, and $j$ = 3, 6, 7) are zero.
Then, $M_{ij}$ ($i, j$ = 1, 2, 4, 5) become the elements of the neutral scalar Higgs boson mass matrix while $M_{ij}$ ($i, j$ = 3, 6, 7) become those of the neutral pseudoscalar Higgs boson one.
In general, because the complex phases are nonzero in spontaneous CP violation scenario, $M_{ij}$ ($i$ = 1, 2, 4, 5, and $j$ = 3, 6, 7) are nonvanishing.

Consider, for example, $M_{13}$ and $M_{23}$.
$M_{13}$ describes the mixing between $S_1$ and $A$. 
Similarly, $M_{23}$ describes the mixing between $S_2$ and $A$. 
We can reexpress $M_{13}$ by substituting the minimum condition for $k_{11}$ into it as
\[
        -{8 \over 3} \lambda_1 k_{12} x_1 x_2 \sin (2 \phi - \phi_1 - \phi_2) \cos\beta \ ,
\]
and similarly for $M_{23}$ by exchanging $\cos \beta \leftrightarrow \sin\beta$.
One can easily notice that the source of nonzero $M_{13}$ or $M_{23}$ is $- 4 \lambda_1 H_1^T \epsilon H_2 N_1^* N_2^*$ +  H.c. terms in $V_F$ of the tree-level Higgs potential.
Therefore, the situation is different from the (M+1)SSM, where the scalar-pseudoscalar mixings do not occur in any of the two Higgs doublets at the tree-level (M+1)SSM.
We would like to emphasize that in the (M+2)SSM there exists scalar-pseudoscalar mixings even at the tree level by the minimum conditions.
In other words, if the Higgs potential of the (M+2)SSM at the tree level is such that the VEVs of four neutral Higgs fields are complex, the mixings between scalar and pseudoscalar Higgs fields would take place generally, which implies the occurrence of the spontaneous CP violation.

\section{NUMERICAL ANALYSIS}

Now, we have to check if our model is phenomenologically acceptable. 
That is, according to either or not the model possess a significantly large region in its parameter space that is consistent with the present phenomenology, the model becomes either physical or merely mathematical, respectively.
For this purpose, we would like to examine if our model predicts a reasonable mass for the lightest neutral Higgs boson, and if it passes the experimental criterion set by LEP2 data. 
These jobs are performed numerically by using Monte Carlo method. 

The free parameters we have in the tree-level Higgs sector of the (M+2)SSM with spontaneous CP violation are $\phi$, $\phi_1$, $\phi_2$, $\tan \beta$, $x_1$, $x_2$, $A_{\lambda_1}$, $\lambda_1$, $k_{12}$, $m_{\tilde t}^2$ and $m_{\tilde b}^2$.
For our numerical analysis, the ranges for the parameters are set as follows:
$0  < \phi, \phi_1, \phi_2 \le \pi$, $2 \le \tan \beta  \le  40$, $0  < x_1, x_2, A_{\lambda_1}  \le 500$ \ GeV, $0 < \lambda_1 \le 0.87$, and $0 < k_{12} \le 0.63$.
These ranges define the extent of the parameter space we will explore.
We fix $m_{\tilde t}^2$ and $m_{\tilde b}^2$ in the effective potential at 1000 GeV.

Let us first estimate the neutral Higgs boson masses by diagonalizing the $7\times7$ symmetric mass matrix of the neutral Higgs bosons. 
The seven eigenvalues are sorted such that $m_{h_i} < m_{h_j}$ for $i<j$.
We scan $10^6$  points in the relevant paramter space and numerically evaluate 
$m_{h_i}$ ($i$ =1 to 7). 
For each point, we also evaluate the mass of charged Higgs boson. 
We find that the masses of neutral Higgs bosons as well as the charged Higgs boson mass are positive for the most part of the parameter space of the (M+2)SSM that we consider. 
Consequently, we note that no serious reduction occurs in the considered parameter space by either the spontaneous electroweak symmetry breaking or the minimum conditions for the vacuum.

Next, we consider the production of a neutral Higgs boson in $e^+e^-$ collisions.
For the center of mass energy of LEP2, the neutral Higgs bosons are dominantly produced through the Higgs-strahlung process $e^+e^- \rightarrow Z^* \rightarrow Z h_i$ ($i$ = 1 to 7) and the Higgs-pair production process  $e^+e^- \rightarrow Z^* \rightarrow h_i h_j$ ($i$, $j$ = 1 to 7, where $i \neq j$).
The relevant coupling coefficients for the these processes are
\begin{eqnarray}
        G_{Z Z h_i}^2 & = &\mbox{} {g_2^2 m_Z^2 \over  4 \cos^2 \theta_W} 
                        (\cos \beta O_{i1} + \sin \beta O_{i 2})^2 \ , \cr
        G_{Z h_i h_j}^2 & = & {g_2^2 \over 4 \cos^2 \theta_W}
        [ \{O_{j 3} (\cos \beta O_{i 2} - \sin\beta O_{i 1})
        - O_{i 3} (\cos \beta O_{j 2} - \sin\beta O_{j 1}) \}^2 \cr
        & &\mbox{} + \{O_{i 1} O_{j 1} + O_{i 2} O_{j 2} + O_{i 3} O_{i 3} \}^2 ] \ ,
\end{eqnarray}
where $\theta_W$ is the Weinberg mixing angle, and $O_{ij}$ is the elements of the orthogonal transformation matrix that diagonalizes the mass matrix of the neutral Higgs bosons.
Note that the expression for the coupling coefficient in the Higgs pair production is different from that of the MSSM. 
The  production cross sections of the Higgs bosons receive the effects of spontaneous CP violation and the radiative effects of the degenerate scalar top quark masses through the above coupling coefficients via $O_{ij}$.

Although candidates of detecting a Higgs boson has been suggested by LEP2 [18], no concrete evidence is available yet; thus, the parameter space of the (M+2)SSM may be constrained experimentally if we assume the discovery limit of 0.1 pb at LEP 2 with $\sqrt{s}$ = 200 GeV.
Let us denote $\sigma_t$ the total cross section for the neutral Higgs productions via the above two processes. 
The total cross section is evidently dependent on the neutral Higgs boson masses.
For a given set of parameter values, we evaluate $m_{h_i}$ numerically in order to insert into the formula of $\sigma_t$. 
Then we evaluate $\sigma_t$ numerically.
We carry out the analysis for the whole parameter space we have defined, by choosing randomly a total of $10^6$ points in the space.
Finally, we select those points which yield $\sigma_t<$ 0.1 pb. They are consistent with experimental constraints and thus establish a subset of parameter space that are phenomenologically acceptable.

We find that there is an indirect lower limit of 110 GeV for the mass of charged Higgs boson by considering in this way the failure of the neutral Higgs boson production at LEP2 with $\sqrt{s}$ = 200 GeV. 
This might be considered as its phenenomelogical lower bound in our model. 
Since the two neutral Higgs singlets in our model distinguish it most distinctively from (M+1)SSM, it is worthwhile to study the behaviors of their VEVs.
In Fig. 1, we show the points that are accepted by the above procedure in ($x_1$, $x_2$) plane.
It is interesting to notice in Fig. 1 that both $x_1$ and $x_2$ cannot simultaneously be small: Roughly, they cannot simultaneously be smaller than $\sim$ 20 GeV.

In Fig. 2(a), we plot $m_{h_1}$ (solid curve) and $m_{h_2}$ (dashed curve) as a function of $\tan \beta$, for $\phi = \pi/10$, $\phi_1 = 2 \pi/3$, $\phi_2 = \pi/20$, $x_1 = 200$ GeV, $x_2 = 400$ GeV, $A_{\lambda_1} = 250$ GeV, $\lambda_1 = 0.1$, $k_{12} = - 0.5$, and $m_{\tilde t} = m_{\tilde b} = 1000$ GeV.
In the case of Fig. 2(a), the mass of the charged Higgs boson is estimated to be between 287 and 962 GeV and $\sigma_t$  between 0.1 and 0.4 fb as $\tan \beta$ goes from 2 to 40.
In Fig 2(b), we repeat the same calculation as Fig 2(a), except for $x_1 = 400$ GeV and $x_2 = 200$ GeV.
In the case of Fig. 2(b), we have $230 \le m_{H^+} \le 755$ GeV and $8.6 \le \sigma_t \le 53.6$ fb for $2 \le \tan \beta \le 40$.
For larger $\tan \beta$, $\sigma_t$ becomes smaller and $m_{H^+}$ larger.
Those parameter values are consistent with the phenomenological constraints of LEP2 at $\sqrt{s} = 200$ GeV.
Figures 2 display that two neutral Higgs bosons $h_i$ ($i$ = 1, 2) are relatively heavy for the allowed range of $2 \le \tan \beta \le 40$. 
Those results are quite dissimilar to the case of the (M+1)SSM.
In the case of the (M+1)SSM, the sum of two lightest neutral Higgs boson masses is smaller than 100 GeV and the charged Higgs boson mass is smaller than 110 GeV [13].

\section{CONCLUSIONS}

It is known that in supersymmetric models with at least two Higgs doublets CP symmetry may be violated through the mixing between the scalar Higgs boson and the pseudoscalar one in the neutral Higgs sector.
For the tree-level Higgs potential of the (M+1)SSM, spontaneous CP violation cannot occur in its neutral Higgs sector since the CP violating vacuum is not stable and thus the mass of the lightest neutral Higgs boson is negative for a reasonable part of the parameter space.

Within the context of the (M+1)SSM, spontaneous CP violation scenario is hardly improved when the 1-loop effective potential due to the degenerate scalar top quark masses is taken into account.
We have investigated whether or not the above situation might be effectively improved by considering the (M+2)SSM with the 1-loop effective potential due to the degenerate scalar quark masses for the third generation.
The most general form of the tree-level Higgs potential is constructed, where the interaction terms between the two Higgs singlet fields are included.
We also take into account the radiative effects due to the degenerate scalar quark masses of the third generation to the tree-level Higgs potential.

The Higgs potential is supposed to develop complex VEVs for the four neutral components of the Higgs multiplets, in order to accommodate spontaneous CP violation. 
From the complex VEVs, three complex phases appear as independent free parameters of the model.
With respect to these complex phases, three minimum conditions for the Higgs potential are calculated to reduce the number of free parameters by three. 
Then, the mass matrix for the seven neutral Higgs bosons are calculated. 
In the (M+2)SSM, the matrix elements ($M_{13}, M_{23}$) for $S_1$-$A$ mixing and $S_2$-$A$ one would not vanish for spontaneous CP violation scenario.
The presence of the mixing elements is found to be essential for spontaneous CP violation.

By assuming the discovery limit of 0.1  pb at LEP2 with $\sqrt{s} = 200$ GeV, we explore the parameter space of the Higgs sector of the (M+2)SSM, if the lightest neutral Higgs boson might have escaped from detection at LEP2.
Our model is numerically analyzed if it is phenomenologically acceptable. 
In spontaneous CP violation scenario, the mass of the lightest neutral Higgs boson is evaluated to be as large as about 80 GeV for some parameter space.
The charged Higgs boson has 110 GeV as the phenomenological lower bound on its mass in our model. 
Concludingly, we have found that spontaneous CP violation in the (M+2)SSM can be accommodated for the Higgs potential including radiative effects due to the degenerate scalar quark masses.

\vskip 0.3 in

\noindent
{\large {\bf ACKNOWLEDGMENTS}}             
\vskip 0.1 in

\noindent
This work was supported by Korea Research Foundation Grant (KRF-2001-050-D00005).

\vskip 0.1 in
\vfil\eject

{\large {\bf APPENDIX: HIGGS BOSON MASS MATRIX}}

At the tree level the elements of the neutral Higgs boson mass matrix are expressed as follows:
\begin{eqnarray}
 M_{11} & = & (m_Z \cos \beta)^2 + \lambda_1 A_{\lambda_1} 
\left \{x_1 \cos \left(\phi - {\phi_2 \over 3} \right) 
+ x_2 \cos \left(\phi - {\phi_1 \over 3}\right) \right \} \tan \beta  \ , \cr
 & &\mbox{}  + {4  \over 3}  \lambda_1 k_{11}  
\left \{x_1^2  \cos \left(\phi  - \phi_1 
-{\phi_2 \over 3}\right) + x_2^2 \cos \left(\phi - {\phi_1 \over 3} - \phi_2
\right) \right \} \tan \beta        \cr
 & &\mbox{} + {4 \over 3}  \lambda_1 k_{12} x_1 x_2 \cos  (2 \phi - \phi_1 - \phi_2) \tan \beta  \ , \cr
 M_{12} & = & 2 v^2 \lambda_1^2
\sin 2 \beta - {m_Z^2 \over 2} \sin 2 \beta - \lambda_1 A_{\lambda_1}
\left \{x_1 \cos \left(\phi - {\phi_2 \over 3} \right) 
+ x_2 \cos \left(\phi - {\phi_1 \over 3}\right) \right \} \cr
 & &\mbox{} - {4 \over 3} \lambda_1 k_{11} \left \{x_1^2 \cos \left(\phi - \phi_1 -{\phi_2 \over 3}\right) + x_2^2 \cos \left(\phi - {\phi_1 \over 3} 
 - \phi_2 \right) \right \}    \cr
 & &\mbox{} - {4 \over 3} \lambda_1 k_{12} x_1  x_2 \cos ( 2 \phi - \phi_1 - \phi_2)   \ , \cr
  M_{13} & = & \lambda_1 A_{\lambda_1}
\left \{x_1 \sin \left(\phi - {\phi_2 \over 3} \right) + x_2 \sin
\left(\phi - {\phi_1 \over 3} \right) \right \} \cos \beta \cr
 & &\mbox{} + {4 \over 3} \lambda_1 k_{11} 
\left \{x_1^2 \sin \left(\phi - \phi_1 -{\phi_2 \over 3} \right) + x_2^2 \sin \left(\phi - {\phi_1 \over 3} - \phi_2 \right) \right 
\} \cos \beta    \cr
 & &\mbox{} + {4 \over 3} \lambda_1 k_{12} x_1 x_2 
 \sin ( 2 \phi - \phi_1 - \phi_2) \cos \beta  \ , \cr
  M_{14} & = & 2 v  \lambda_1^2 \left \{x_1 + x_2  \cos \left({\phi_1 -\phi_2 \over 3} \right) \right \}\cos \beta
 - v \lambda_1 A_{\lambda_1} \cos \left (\phi - {\phi_2 \over 3} \right)\sin \beta \cr
 & &\mbox{} - {8 \over 3} v \lambda_1  k_{11} x_1 
\cos \left( \phi - \phi_1  - {\phi_2 \over 3}\right)\sin \beta - {4 \over 3} v \lambda_1  k_{12} x_2 \cos (2 \phi  - \phi_1 - \phi_2) \sin \beta   \ , \cr
  M_{15} & = & 2 v  \lambda_1^2 \left \{x_2 + x_1  \cos \left({\phi_1 -\phi_2 \over 3} \right) \right \}
 \cos \beta - v \lambda_1 A_{\lambda_1} \cos \left (\phi - {\phi_1 \over 3} \right)\sin \beta \cr
 & &\mbox{} - {8 \over 3} v \lambda_1 k_{11} x_2 \cos \left( \phi - {\phi_1 \over 3} - \phi_2 \right)\sin \beta
 - {4 \over 3} v \lambda_1  k_{12} x_1 \cos (2 \phi  - \phi_1 - \phi_2 ) \sin  \beta \ , \cr
 M_{16} & = & v  \lambda_1 A_{\lambda_1} \sin \left  ( \phi - {\phi_2  \over 3}\right) \sin \beta - {8 \over 3} v \lambda_1 k_{11} x_1 \sin \left(\phi - \phi_1 - {\phi_2 \over 3} \right) \sin \beta    \cr
 & &\mbox{} - {4 \over 3} v \lambda_1 k_{12} x_2 \sin (2 \phi - \phi_1 - \phi_2) \sin \beta - 2 v \lambda_1^2 x_2 \sin \left ( {\phi_1 - \phi_2 \over 3} \right) \cos \beta \ , \cr
 M_{17} & = & v  \lambda_1 A_{\lambda_1} \sin \left  ( \phi - {\phi_1  \over 3}\right) \sin \beta - {8 \over 3} v \lambda_1 k_{11} x_2 \sin \left(\phi - {\phi_1 \over 3} - \phi_2 \right) \sin \beta    \cr
 & &\mbox{} - {4 \over 3} v \lambda_1 k_{12} x_1 \sin (2 \phi - \phi_1 - \phi_2) \sin \beta + 2 v \lambda_1^2 x_1 \sin \left ( {\phi_1 - \phi_2 \over 3} \right) \cos \beta \ , \cr
 M_{22} & = & M_{11} (\cos \beta \rightarrow \sin \beta, \ \tan  \beta \rightarrow \cot \beta)  \ , \cr
 M_{23} & = & M_{13} (\cos \beta \rightarrow \sin \beta)   \ , \cr
 M_{24} & = & M_{14}(\cos \beta \rightarrow  \sin \beta, \ \sin \beta \rightarrow  \cos \beta) \ , \cr
 M_{25} & = & M_{15}(\cos \beta \rightarrow  \sin \beta, \ \sin \beta \rightarrow  \cos \beta) \ , \cr
 M_{26} & = & M_{16}(\cos \beta \rightarrow  \sin \beta, \ \sin \beta \rightarrow  \cos \beta) \ , \cr
 M_{27} & = & M_{17}(\cos \beta \rightarrow  \sin \beta, \ \sin \beta \rightarrow  \cos \beta) \ , \cr
 M_{33} & = & {\lambda_1 A_{\lambda_1} x_1 \over 4 \sin 2 \beta}
\left \{\cos \left (4 \beta  - \phi + {\phi_2 \over  3} \right ) - \cos  \left (4 \beta - 2 \phi + {\phi_1 \over 3} + {2 \over 3} \phi_2 \right) \right \} \cr
 & &\mbox{} + {\lambda_1 A_{\lambda_1} x_2 \over 4 \sin 2 \beta}
 \left \{\cos \left (4 \beta  - \phi + {\phi_1 \over  3} \right ) - \cos  \left (4 \beta - 2 \phi + {2 \over 3} \phi_1 + {\phi_2 \over 3}  \right) \right
 \} \cr
 & &\mbox{} + {2 \lambda_1 A_{\lambda_1} \over \sin 2 \beta} 
\left \{x_1 \cos \left (\phi - {\phi_2 \over 3} \right )
 + x_2 \cos \left (\phi - {\phi_1 \over 3} \right) \right \} \cr
 & &\mbox{} + {\lambda_1 k_{11} x_1^2 \over 3 \sin 2 \beta} \left \{\cos \left(4 \beta - \phi + \phi_1 + {\phi_2 \over 3} \right )
 - \cos \left (4 \beta - 2 \phi + {4 \over 3} \phi_1 + {2 \over 3} \phi_2 \right) \right \}  \cr
 & &\mbox{} + {\lambda_1 k_{11} x_2^2 \over 3 \sin 2 \beta} \left
\{\cos \left(4 \beta - \phi + {\phi_1 \over 3} + \phi_2 \right )
 - \cos \left (4 \beta - 2 \phi + {2 \over 3} \phi_1 + {4 \over 3} \phi_2 \right) \right \}  \cr
 & &\mbox{} + {8 \lambda_1 k_{11} \over 3 \sin 2 \beta} \left
 \{x_1^2 \cos \left (\phi - \phi_1 - {\phi_2 \over 3} \right )
 + x_2^2 \cos \left (\phi - {\phi_1 \over 3} - \phi_2 \right) \right \} \cr
 & &\mbox{}+ { 8 \lambda_1 k_{12} \over 3 \sin 2 \beta} x_1 x_2 \cos (2 \phi - \phi_1 - \phi_2)  \ ,   \cr
 M_{34} & = & v \lambda_1 A_{\lambda_1} \sin \left ( \phi - {\phi_2 \over 3}\right) + {8 \over 3} v \lambda_1 k_{11} x_1 \sin  \left(\phi - \phi_1 - {\phi_2 \over 3}\right)  \cr
 & &\mbox{} + {4 \over 3} v \lambda_1 k_{12} x_2 \sin ( 2 \phi - \phi_1 - \phi_2)  \  ,   \cr
 M_{35} & = &  M_{34} (\phi_1 \rightarrow  \phi_2, \ \phi_2 \rightarrow  \phi_1, \ x_1 \rightarrow x_2, \ x_2 \rightarrow x_1) \  ,  \cr
 M_{36} & = & v \lambda_1 A_{\lambda_1} \cos \left ( \phi - {\phi_2 \over 3}\right) - {8 \over 3} v \lambda_1 k_{11} x_1 \cos \left(\phi -  \phi_1 - {\phi_2 \over 3}\right)  \cr
 & &\mbox{} - {4 \over 3} v \lambda_1 k_{12} x_2 \cos ( 2 \phi - \phi_1 - \phi_2)  \  ,   \cr
 M_{37} & = &  M_{36} (\phi_1  \rightarrow \phi_2, \ \phi_2 \rightarrow  \phi_1, \ x_1 \rightarrow x_2, \ x_2 \rightarrow x_1)   \  ,  \cr
 M_{44} & = & {40 \over 9} k_{11}^2 x_1^2 - k_{11} A_{k_{11}} x_1 \cos (\phi_1)
 + {8 \over  9 x_1} k_{11}  k_{12} x_2 (3  x_1^2 - x_2^2)  \cos \left ({\phi_1  - \phi_2 \over 3} \right)   \cr
 & &\mbox{} + {4 \over 3} v^2 \lambda_1 k_{11} \cos \left (2  \beta - {3 \phi \over 2} + {7 \phi_1 \over 6} + {\phi_2 \over 2} \right )
 \sin \left ({\phi \over 2} - {\phi_1 \over 6} - {\phi_2 \over 6} \right )  \cr
 & &\mbox{} - {v^2 \over x_1} \lambda_1^2  x_2 \cos \left ( {\phi_1 - \phi_2  \over 3} \right)
 + {v^2 \lambda_1 \over 2 x_1} A_{\lambda_1} \cos \left (\phi - {\phi_2 \over 3} \right) \sin 2 \beta     \cr
 & &\mbox{} + {k_{12}  A_{k_{12}} \over 3 x_1}  x_2^2 \cos \left (  \phi_1 + 2  \phi_2 \over 3 \right)
 + {2 v^2 \over 3 x_1} \lambda_1 k_{12} x_2 \cos \left (2 \phi - \phi_1 - \phi_2 \right ) \sin 2 \beta     \  ,  \cr
 M_{45} & = & v^2 \lambda_1^2 \cos \left ({\phi_1 - \phi_2 \over 3} \right)
 + {8 \over 3} k_{11} k_{12} (x_1^2 + x_2^2) \cos \left ({\phi_1 - \phi_2 \over 3}\right ) \cr
 & &\mbox{}+ {8 \over 3}  k_{11}^2 x_1 x_2 \cos  \left ({2 \phi_1 -  2 \phi_2 \over 3} \right ) - {2 \over 3} v^2 \lambda_1 k_{12} \cos (2 \phi - \phi_1 - \phi_2) \sin 2 \beta    \cr
  & &\mbox{} + {16 \over 9} k_{12}^2 x_1 x_2 - {2 \over 3} k_{12} A_{k_{12}} \left \{x_1 \cos \left ({2 \phi_1 + \phi_2 \over 3} \right ) + x_2 \cos \left({\phi_1 + 2 \phi_2 \over 3} \right) \right \} \ , \cr
 M_{46} & = & 2 k_{11} A_{k_{11}} x_1 \sin (\phi_1) - {4 \over 3} v^2 \lambda_1 k_{11} \sin \left (\phi - \phi_1 - {\phi_2 \over 3} \right )\sin 2 \beta \cr
 & &\mbox{} - {4  \over 3} k_{11}^2  x_2^2 \sin \left  ({2 \phi_1 -  2 \phi_2 \over  3} \right ) - {16 \over 9} k_{11} k_{12} x_1 x_2 \sin \left ({\phi_1 - \phi_2 \over 3} \right) \cr
 & &\mbox{} + {2 \over 3}  k_{12} A_{k_{12}} x_2 \sin \left  ({2 \phi_1 + \phi_2 \over 3}\right ) \ , \cr
 M_{47} & = & v^2 \lambda_1^2 \sin \left ({\phi_1 - \phi_2 \over 3} \right) + {8 \over 9} k_{11} k_{12} ( 3 x_1^2 + x_2^2) \sin \left ({\phi_1 - \phi_2 \over 3} \right ) \cr
 & &\mbox{} +  {8 \over 3}  k_{11}^2 x_1 x_2  \sin \left  ({2 \phi_1 -  2 \phi_2 \over 3}\right ) - {2 \over 3} v^2 \lambda_1 k_{12} \sin (2 \phi - \phi_1 - \phi_2) \sin 2 \beta  \cr
 & &\mbox{} + {2 \over 3} k_{12}  A_{k_{12}} \left \{x_1 \sin \left ({2 \phi_1  + \phi_2 \over 3} \right ) + x_2 \sin \left({\phi_1 + 2 \phi_2 \over 3} \right) \right \} \ , \cr
 M_{55} & =  & M_{44}(\phi_1 \rightarrow  \phi_2, \  \phi_2 \rightarrow \phi_1,  \ x_1 \rightarrow x_2, \ x_2 \rightarrow x_1) \  ,  \cr
 M_{56} & = &  M_{47} (\phi_1 \rightarrow  \phi_2, \ \phi_2 \rightarrow  \phi_1, \ x_1 \rightarrow x_2, \ x_2 \rightarrow x_1) \ ,  \cr
 M_{57} & = &  M_{46} (\phi_1 \rightarrow  \phi_2, \ \phi_2 \rightarrow  \phi_1, \ x_1 \rightarrow x_2, \ x_2 \rightarrow x_1) \ ,  \cr
 M_{66} & = & 3 k_{11} A_{k_{11}} x_1 \cos  (\phi_1) - {8 \over 9 x_1} k_{11} k_{12} x_2 (x_1^2 + x_2^2) \cos \left ({\phi_1 - \phi_2 \over 3} \right)    \cr
 & &\mbox{} - {8  \over 3} k_{11}^2  x_2^2 \cos \left  ({2 \phi_1 -  2 \phi_2 \over 3} \right ) - {v^2 \over x_1} \lambda_1^2  x_2 \cos \left ({\phi_1 - \phi_2 \over  3} \right)  \cr
 & &\mbox{} + {v^2 \over  2 x_1} \lambda_1 A_{\lambda_1}  \cos \left (\phi -  {\phi_2 \over 3} \right ) \sin 2 \beta + {2 v^2 \over 3 x_1} \lambda_1 k_{12} x_2 \cos (2  \phi - \phi_1 - \phi_2)\sin 2 \beta  \cr
  & &\mbox{} + {2 \over  3 } v^2 \lambda_1  k_{11} \sin \left (2  \beta - 2 \phi  + {4 \phi_1 \over 3} + {2 \phi_2 \over 3}  \right ) \cr
  & &\mbox{} + {k_{12} \over 3 x_1} A_{k_{12}} x_2 \left \{ 4 x_1 \cos \left ({2 \phi_1 + \phi_2 \over 3} \right )  + x_2 \cos \left ({\phi_1 +  2 \phi_2 \over 3} \right)\right \}   \cr
  & &\mbox{} + {2 \over 3}  v^2 \lambda_1 k_{11} \left \{2 \sin  \left(2 \beta + \phi - \phi_1 - {\phi_2 \over 3} \right) + \sin \left (2 \beta - \phi +  \phi_1 + {\phi_2 \over 3} \right ) \right \}   \   ,  \cr
 M_{67} & = & v^2 \lambda_1^2 \cos \left ({\phi_1 - \phi_2 \over 3} \right )
 + {8 \over 9} k_{11} k_{12} (x_1^2 + x_2^2) \cos \left ({\phi_1 - \phi_2 \over 3} \right ) \cr
 & &\mbox{} + {8 \over 3} k_{11}^2  x_1 x_2 \cos \left ({2 \phi_1  - 2 \phi_2 \over 3} \right ) + {2 \over 3} v^2 \lambda_1 k_{12} \cos ( 2 \phi - \phi_1 - \phi_2) \sin 2 \beta \cr
 & &\mbox{} + {2 \over 3} k_{12} A_{k_{12}}  \left \{x_1 \cos \left ({2 \phi_1 + \phi_2 \over 3} \right ) + x_2 \cos \left ({\phi_1 + 2 \phi_2 \over 3} \right )  \right \}    \  ,  \cr
 M_{77} & = &  M_{66} (\phi_1 \rightarrow  \phi_2, \ \phi_2 \rightarrow  \phi_1, \ x_1 \rightarrow x_2, \ x_2 \rightarrow x_1) \ . \nonumber
\end{eqnarray}


\vfil\eject

{\bf Figure Captions}
\vskip 0.3 in
\noindent
Fig. 1 : The points in ($x_1$,  $x_2$)-plane that satisfy $\sigma_t<$  0.1 pb are plotted for 0 $ < x_1, x_2 \le $ 100 GeV by random number generation method.The ranges of other parameters are: 0  $< \phi, \phi_1, \phi_2 \le \pi$, 2 $\le \tan \beta  \le $ 40, 0 $ < A_{\lambda_1} \le$ 500 GeV, 0 $< \lambda_1 \le$ 0.87, and $k_{12} <$ 0.63.

\vskip 0.2 in
\noindent
Fig. 2(a) : $m_{h_1}$ (solid curve) and $m_{h_2}$ (dashed curve) are plotted as a function of $\tan \beta$, for $\phi = \pi/10$, $\phi_1 = 2 \pi/3$, $\phi_2 = \pi/20$, $x_1 = 200$ GeV, $x_2 = 400$ GeV, $A_{\lambda_1} = 250$ GeV, $\lambda_1 = 0.1$, $k_{12} = - 0.5$, and $m_{\tilde t} = m_{\tilde b} = 1000$ GeV.
The plotted results satisfy $\sigma_t<$ 0.1 pb.

\vskip 0.2 in
\noindent
Fig. 2(b) : The same as Fig 2(a), except for $x_1 = 400$ GeV and $x_2 = 200$ GeV.

\vfil\eject

\setcounter{figure}{0}
\def\figurename{}{}%
\renewcommand\thefigure{Fig. 1}
\begin{figure}[t]
\epsfxsize=13cm
\hspace*{2.cm}
\epsffile{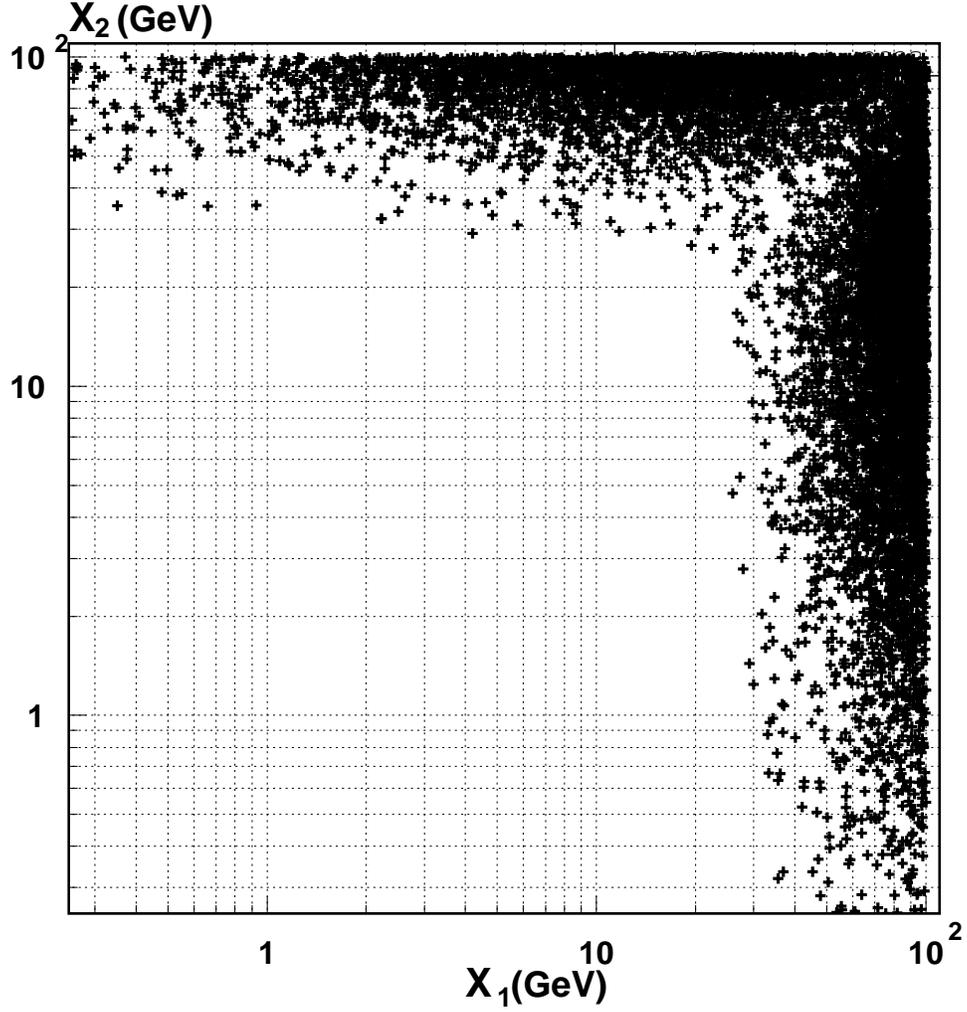}
\caption[plot]{The points in ($x_1$,  $x_2$)-plane that satisfy $\sigma_t<$  0.1 pb are plotted for 0 $ < x_1, x_2 \le $ 100 GeV by random number generation method. 
The ranges of other parameters are: 0  $< \phi, \phi_1, \phi_2 \le \pi$, 2 $\le \tan \beta  \le $ 40, 0 $ < A_{\lambda_1} \le$ 500 GeV, 0 $< \lambda_1 \le$ 0.87, and $k_{12} <$ 0.63.}
\end{figure}

\renewcommand\thefigure{Fig. 2a}
\begin{figure}[t]
\epsfxsize=13cm
\hspace*{2.cm}
\epsffile{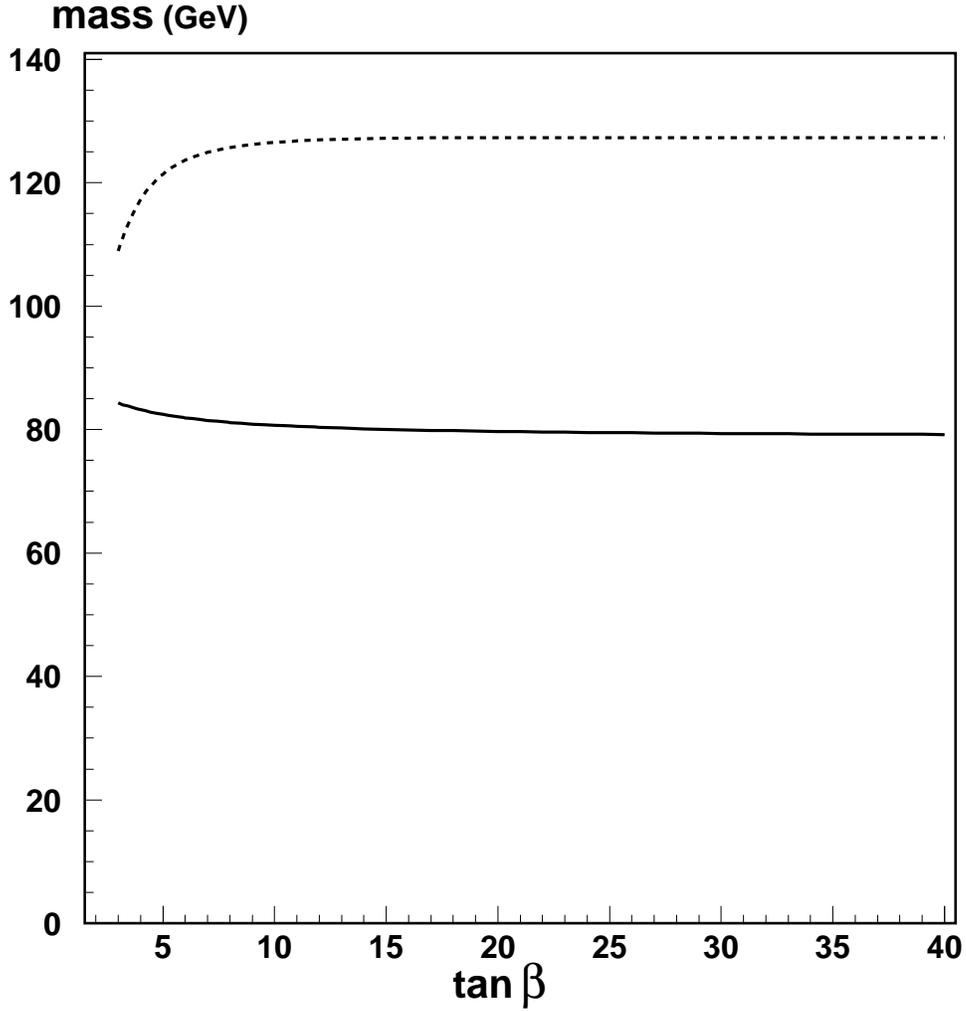}
\caption[plot]{$m_{h_1}$ (solid curve) and $m_{h_2}$ (dashed curve) are plotted as a function of $\tan \beta$, for $\phi = \pi/10$, $\phi_1 = 2 \pi/3$, $\phi_2 = \pi/20$, $x_1 = 200$ GeV, $x_2 = 400$ GeV, $A_{\lambda_1} = 250$ GeV, $\lambda_1 = 0.1$, $k_{12} = - 0.5$, and $m_{\tilde t} = m_{\tilde b} = 1000$ GeV.
The plotted results satisfy $\sigma_t<$ 0.1 pb.}
\end{figure}

\renewcommand\thefigure{Fig. 2b}
\begin{figure}[t]
\epsfxsize=13cm
\hspace*{2.cm}
\epsffile{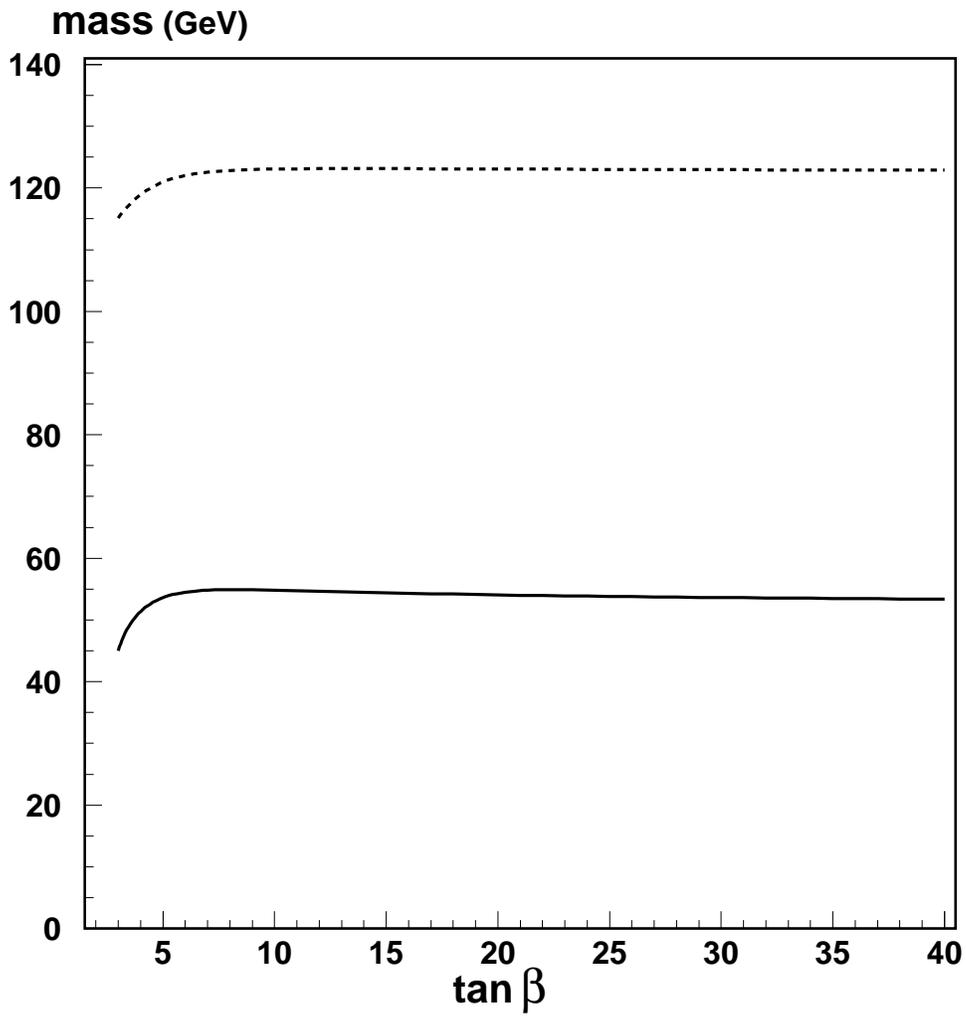}
\caption[plot]{The same as Fig 2(a), except for $x_1 = 400$ GeV and $x_2 = 200$ GeV}
\end{figure}

\end{document}